\documentclass[conference]{IEEEtran}
\IEEEoverridecommandlockouts
\ifCLASSINFOpdf
\else
\fi

\ifCLASSOPTIONcompsoc
  \usepackage[caption=false,font=normalsize,labelfont=sf,textfont=sf]{subfig}
\else
  \usepackage[caption=false,font=footnotesize]{subfig}
\fi

\usepackage{booktabs}
\usepackage{paralist}
\usepackage[group-separator={,}, group-four-digits, per-mode=symbol, binary-units]{siunitx}
\usepackage{color}
\usepackage{xspace}

\usepackage{graphicx}
\usepackage{color, colortbl}
\usepackage{amsfonts}
\usepackage{amsmath}

\usepackage{threeparttable, tablefootnote}

\newcommand{\ie}{i.\@\,e.\@\xspace}
\newcommand{\eg}{e.\@\,g.\@\xspace}
\newcommand{\etal}{et~al.\@\xspace}
\newcommand{\tline}[2][c]{%
\begin{tabular}[#1]{@{}c@{}}#2\end{tabular}}

\begin{document}

\title{Intelligent Reflecting Surface-Assisted Wireless Key Generation for Low-Entropy Environments}

\author{\IEEEauthorblockN{
Paul Staat\textsuperscript{1},
Harald Elders-Boll\textsuperscript{2},
Markus Heinrichs\textsuperscript{2},\\ 
Rainer Kronberger\textsuperscript{2},
Christian Zenger\textsuperscript{3,4}, and
Christof Paar\textsuperscript{1}}\\

\IEEEauthorblockA{\small \textsuperscript{1}Max Planck Institute for Security and Privacy, Bochum, Germany}
\IEEEauthorblockA{\small \textsuperscript{2}TH Köln – University of Applied Sciences, Cologne, Germany}
\IEEEauthorblockA{\small \textsuperscript{3}PHYSEC GmbH, Bochum, Germany}
\IEEEauthorblockA{\small \textsuperscript{4}Ruhr University Bochum, Germany}
%\IEEEauthorblockA{\small Corresponding author E-Mail: paul.staat@csp.mpg.de\textsuperscript{1}}
}

% make the title area
\maketitle

\begin{abstract}
Physical layer key generation is a promising candidate for cryptographic key establishment between two wireless communication parties. It offers information-theoretic security and is an attractive alternative to public-key techniques. Here, the inherent randomness of wireless radio channels is used as a shared entropy source to generate cryptographic key material. However, practical implementations often suffer from static channel conditions which exhibit a limited amount of randomness. In the past, considerable research efforts have been made to address this fundamental limitation. However, current solutions are not generic or require dedicated hardware extensions such as reconfigurable antennas. In this paper, we propose a novel wireless key generation architecture based on randomized channel responses from an intelligent reflecting surface~(IRS). Due to its passive nature, a cooperative IRS is well-suited to provide randomness for conventional resource-constrained radios. We conduct the first practical studies to successfully demonstrate IRS-based physical-layer key generation with an OFDM system. In a static environment, using a single subcarrier only, our IRS-assisted prototype system achieves a key generation rate~(KGR) of \SI{97.39}{bps} with \SI{6.5}{\percent} key disagreement rate~(KDR) after quantization, while passing standard randomness tests.
\end{abstract}

\IEEEpeerreviewmaketitle

\section{Introduction}

Pervasive wireless networks such as used for IoT systems, are a central aspect of today's connected world and must be thoroughly protected against attacks. In this context, symmetric encryption schemes such as AES play an important role for providing  confidentiality as well as message integrity and  authentication. However, this requires a key exchange mechanism.

As an alternative to classical public-key techniques for secret sharing over public channels such as the wireless radio channel, significant research efforts have been devoted to primitives from the realm of physical layer security~(PLS)~\cite{hamamrehClassificationsApplicationsPhysical22}. In particular, channel reciprocity-based key generation~(CRKG) leverages randomly behaving wireless radio channels to establish shared cryptographic keys to achieve information-theoretic security. 

Despite the large body of work exploring CRKG in recent years~\cite{zhang_key_2016}, the acceptance for real-world deployment is rather low as the performance of CRKG is tied to the wireless channel conditions. For instance, static radio channels can only provide a limited amount of entropy and substantially impede the key generation process: Establishing sufficiently random keys is impractically time-consuming at low key generation rates. This issue affects a wide range of wireless applications with limited mobility, i.e., where channel conditions tend to be static. Examples include warehouses, enclosures, and at night  also many other indoor environments.

The ability to work properly in static environments has been outlined as one of the major challenges for physical-layer key generation~\cite{zhang_key_2016}. Therefore, several previous works investigate approaches to tackle static channel conditions, \eg, reconfigurable antennas, jamming, and beamforming~\cite{aono_wireless_2005, gollakota_physical_2011, zhang_key_2016}. However, none of them is generically applicable to existing CRKG systems as dedicated hardware extensions or special modulation schemes are required. This is a serious hurdle for the wide-range adoption of the technology, especially for low-resource devices such as found in many IoT systems, for which CRKG in general is desirable.

In this paper we pursue a generic solution based on an intelligent reflecting surface~(IRS) to enable CRKG in static environments for arbitrary devices while being compatible with existing CRKG implementations. Here, the IRS serves as an entropy source for the wireless channel, offering an entirely new range of applications for future IRS deployments.

The IRS concept has evolved from research on metamaterials, which are synthetic structures with tailored EM characteristics to realize non-standard wave manipulation capabilities. Cost-effective digitally tunable and flat metamaterial variants paved the way for IRS to become attractive for future communication systems beyond 5G. It has gained significant research interest~\cite{basar_wireless_2019} due to its innovative nature: Adding control to the wireless propagation environment, the IRS enables what is coined \textit{smart radio environments}. The IRS intelligently interacts with radio waves in an entirely passive manner with moderate hardware complexity and low energy consumption.

In the context of this work, we use a cooperative IRS to diminish static channel conditions, assisting CRKG. In particular, we deliberately randomize the wireless channel and generate temporal variation to provide an entropy source which is independent of user terminals.
We implement a prototype system using commodity \mbox{Wi-Fi} transceivers and low-complexity IRS prototypes operating in the~\SI{5}{GHz} frequency range.
A key characteristic of our approach is that we consider the wireless channel to provide an additional degree of freedom when designing a wireless key generation system. To the best of our knowledge, this is the first work to implement a practical CRKG system incorporating an IRS. 
The paper at hand contains the following key contributions:
\begin{compactitem}
    \item We propose physical layer key generation assisted by an IRS to overcome otherwise static propagation environments. Our approach addresses critical real-world requirements of low-resource CRKG systems.

    \item We show that IRS-assisted CRKG can achieve arbitrary and adjustable key generation rates, while ensuring random key material. Further, we introduce a channel oversampling technique to reduce bit mismatch in the generated key material.

    \item We implement a functional proof-of-concept system based on low-cost IRS prototypes and commodity MIMO radio transceivers. 
    We present a comprehensive evaluation based on measurements, showing that IRS-assisted CRKG is practically feasible.

\end{compactitem}

\section{Background}

\subsection{Channel Reciprocity-Based Key Generation}

The fundamental principles of \textit{channel reciprocity} and \textit{spatial decorrelation}~\cite{goldsmith_wireless_2005} allow to exploit the wireless channel as a mutual keying variable, \ie, for CRKG. Secret key agreement over authenticated two-way public channels from dependent random variables has been pioneered by Maurer~\cite{maurer_secret_1993}.
Since then, much work has dealt with practice-oriented wireless CRKG protocols, including prominent examples by Mathur~\etal~\cite{mathur_radio-telepathy_2008}, Patwari~\etal~\cite{mathur_radio-telepathy_2008}, and Aono~\etal~\cite{aono_wireless_2005}. These protocols follow the same rationale: In a first step, the two participants exchange a series of messages to collect channel measurements, \eg, received signal strength~(RSS) or channel state information~(CSI). Then, a quantization stage maps the channel observations to bits, producing correlated bit sequences $K^A$ and $K^B$ at both nodes. An error correcting code is used for information reconciliation, \ie, to combat bit mismatch. Finally, privacy amplification (via hashing) removes information that has leaked during the exchange of helper data for error correction. Important metrics for CRKG systems include the similarity of channel observations at both ends (\eg, mutual information and correlation measures), the key generation rate~(KGR), and the key disagreement rate~(KDR).%, and

Most of the current CRKG schemes gather data from time-variant channels and were designed under the assumption of user movement and random channel variation and in view of statistical channel models such as Rayleigh fading. However, these schemes fail to work with static radio channels. 
Previous solutions to combat static channels come at the cost of significantly increased complexity: For instance, \cite{aono_wireless_2005} leverages an electronically steerable antenna, while \cite{gollakota_physical_2011} describes a jamming-based technique.

\subsection{Intelligent Reflecting Surface}

An IRS, sometimes also referred to  metasurface in the literature, is a man-made planar structure with digitally controllable electromagnetic reflection behavior. The IRS adds control to the propagation of radio waves and thereby can optimize wireless radio channels.  IRS' are sometimes considered a paradigm shift towards \textit{smart radio environments}~\cite{liaskos_novel_2019} and are already in discussion for future communication networks beyond 5G~\cite{basar_wireless_2019}.
 
The IRS has potential for innovation at relatively low hardware complexity. Typically fabricated in microstrip technology on low-cost printed circuit board~(PCB) substrate, the IRS consists of many individually tunable reflector elements. For instance, the IRS controller can adjust the phase shift of reflections across the surface to optimize the signal-to-noise ratio~(SNR) at a receiver~\cite{basar_wireless_2019} or to enhance spatial diversity~\cite{del_hougne_optimally_2019}. Due to the mostly passive and reflecting nature, the IRS does not require active RF chains and is inherently capable of full duplex operation, while being energy efficient.

In the PLS context, previous work with IRS addressed degradation of the eavesdropper's channel, \eg, from exclusion~\cite{liaskos_novel_2019}. For CRKG, \cite{jiSecretKeyGeneration2021} has outlined an IRS-assisted approach to reduce information leakage to the eavesdropper. 

\subsection{System and Adversary Model}
In this work, we consider two legitimate parties Alice and Bob who seek to establish a shared cryptographic key. Alice and Bob deploy a standard time-divison duplex~(TDD) wireless communication protocol, \eg, IEEE 802.11n \mbox{Wi-Fi} with orthogonal frequency-division multiplexing~(OFDM), and obtain CSI as an essential part of their communication. Using their respective CSI data, both parties implement a CRKG procedure. 
Furthermore, we assume that a passive IRS is within reach of Alice and/or Bob and thus can partially control the wireless propagation channel.
The passive eavesdropper Eve can capture messages sent by Alice and Bob and also obtains CSI, representing the effective channels between Eve and the legitimate parties Alice and Bob, respectively. Eve is aware of the key generation procedure of Alice and Bob and thus can derive own key material.

All parties operate in a normally static indoor environment, \ie, the parties do not move, nor do objects in the environment. Hence, the environment does not contribute to temporal variation and thus exhibits a limited amount of randomness. 
We assume the passive IRS to be the only source of channel variation.

\section{IRS-Assisted Key Generation}
\begin{figure}
\centering
\includegraphics[width=0.9\linewidth]{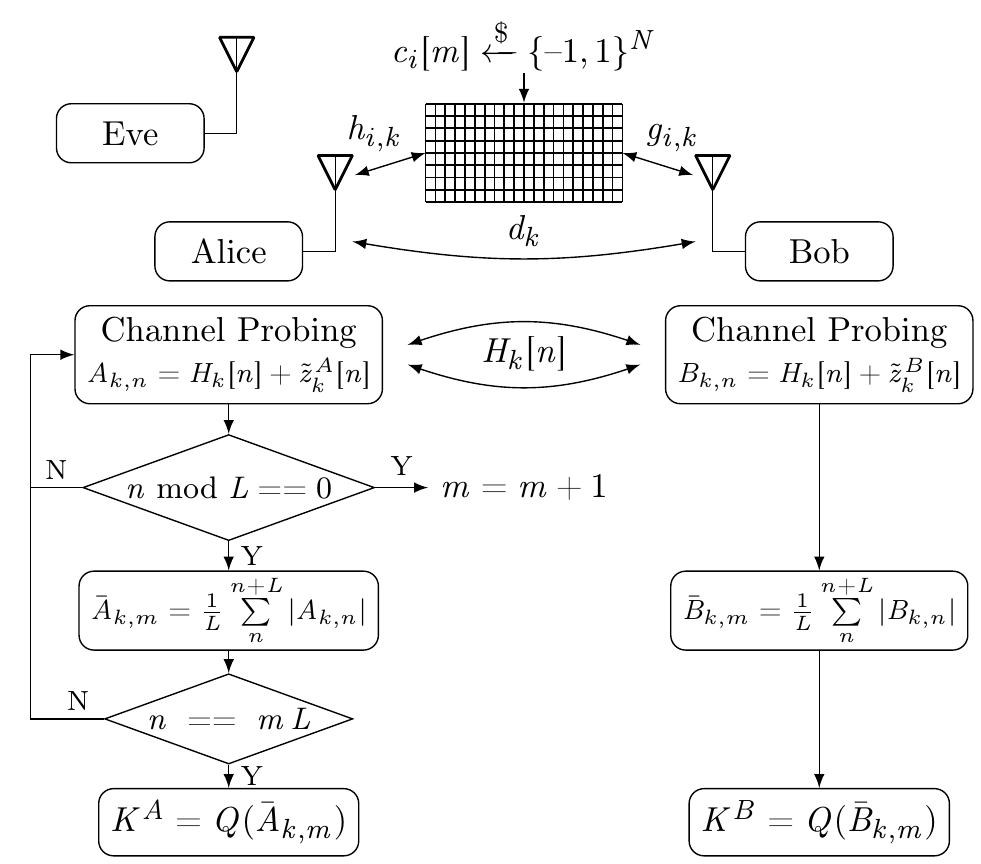}
\caption{Key generation procedure assisted by an IRS.}
\label{fig:block_diag_keygen}
\end{figure}

Our IRS-assisted CRKG scheme follows a standard architecture which includes channel probing, quantization, information reconciliation, and privacy amplification stages. The key difference of our scheme is the channel probing which is performed in conjunction with a channel randomization step from the IRS (see Fig.~\ref{fig:block_diag_keygen}). We review the protocol initiation, channel randomization and probing, synchronization, and quantization steps. We do not elaborate on information reconciliation and privacy amplification since well-known approaches from literature can be used~\cite{bloch_wireless_2008}.

\subsubsection{Initiation}
During protocol initiation, the nodes and the IRS exchange the timing for channel probing, the total number of surface configurations~$m$, and the oversampling factor~$L$. The latter determines the number of bidirectional samples per surface configuration. In addition, the nodes can verify the IRS' channel impact.

\subsubsection{Channel Randomization}
The generalized reflection coefficients for the $N$-element IRS can be expressed as:
\begin{equation}
    \phi_i[n] = \alpha_i e^{j \varphi_i[n]} = c_i[n] \qquad i = 1,...,N
\end{equation}
where $\alpha_i = 1$. In the context of this work, we consider a binary-tunable IRS with 1-bit phase control per element, restricting $\varphi_i \in \{0, \pi\}$. Hence, we can substitute $\phi_i[n]$ with the surface control signal $c_i[n] \in \{-1,1\}^N$. 

Since the prototype system is implemented by a set of commercial off-the-shelf \mbox{Wi-Fi} Network Interface Cards (NICs), we assume that the complex OFDM baseband signal transmitted by Alice is generated by taking the inverse discrete Fourier transform of the complex modulated 
data symbols $X^A_k[n]$ of all $K$, $k= 0,\ldots, K-1$, subcarriers of the $n^{th}$ OFDM symbol. In the time domain, a cyclic prefix, which is assumed to be longer than the channel's maximum delay spread, is then prepended to each OFDM symbol. Then, after time- and frequency synchronization, removal of the cyclic prefix and discrete Fourier transform, the received baseband signal of Bob corresponding to the $k^{th}$ subcarrier of the $n^{th}$~OFDM~symbol in the frequency domain is given by:  
\begin{equation}
\label{eq:tx_rx_h}
    %Y^A_k[n] = \left( \sum_{i=0}^N h_{i,k} e^{j \phi_i[n]} g_{i,k} + d_k \right) x_k[n] + z^A_k[n]
    Y^B_k[n] = \left( \sum_{i=0}^N h_{i,k}\, c_i[n]\, g_{i,k} + d_k \right) X^A_k[n] + z^B_k[n],
    %Y = X \left( \sum_{n=0}^N \phi_n g_n + d \right) + v
    %h(t, c) = \sum_{n=0}^{N} e^{j \phi(c[n])}\ \alpha_n \delta(t-\tau_n) + \sum_{k=0}^{K} \beta_k \delta(t-\tau_k).
\end{equation}
where  $h_{i,k}, g_{i,k}, d_k \in \mathbb{C}$, respectively, are the complex channel gains of the link between Alice and the $i^{th}$ IRS element, Bob and the $i^{th}$ IRS element, the direct link between Alice and Bob for the $k^{th}$ subcarrier, and $z^B_k[n] \sim \mathcal{CN}(0,\sigma^2)$ is additive white Gaussian noise~(AWGN). As all links are reciprocal, the received baseband signal $Y^A_k[n]$ of Alice can be noted in the same manner by exchanging $A$ and $B$ in (\ref{eq:tx_rx_h}). For CRKG, we are interested in the effective channel
\begin{equation}
    H_k[n] = \sum_{i=0}^N h_{i,k}\, c_i[n]\, g_{i,k} + d_k.
    \label{eq:h_eff}
\end{equation}

We assume that the surface configurations $c_i[n]$ are selected randomly. From the central limit theorem, we gain a first insight into the effectiveness of the IRS for channel randomization: If the surface is sufficiently large, \ie, $N \gg 1$, $H_k[n]$ converges to a complex normal distribution with variance linearly scaling with $N$.

\subsubsection{Channel Probing}
During channel probing, Alice and Bob exchange packets in a ping-pong manner. At the receiver side, Alice and Bob use a standard Least-Squares~(LS) channel estimator to obtain CSI:

\begin{gather}
    \hat{H}^A_{k}[n] = \frac{Y^A_k[n]}{X^B_k[n]} = H_k[n] + \frac{z^A_k[n]}{X^B_k[n]} =  H_k[n] + \tilde{z}^A_k[n], \label{eq:channel_probing_alice} \\%H_k[n] + \frac{z_k[n]}{x^A_k[n]} %\tilde{z}^A_k[n]\\
    \hat{H}^B_k[n] = \frac{Y^B_k[n]}{X^A_k[n]} = H_k[n] + \frac{z^B_k[n]}{X^A_k[n]} =  H_k[n] + \tilde{z}^B_k[n].%H_k[n] + \frac{z_k[n]}{x^B_k[n]} %\tilde{z}^B_k[n]
    \label{eq:channel_probing_bob}
\end{gather}

Note that we assume the IRS to be the only source of channel variation. As the IRS switches channel states, Alice and Bob effectively observe a static channel during one IRS configuration period. Thus, averaging over $L$~consecutive samples obtained per IRS configuration can help Alice and Bob to reduce noise components (see Fig.~\ref{fig:block_diag_keygen}).

\subsubsection{Synchronization}

The nodes are either aware of or can alternatively measure the IRS modulation frequency. The latter is possible since the IRS modulates the channel response periodically~(cf. Fig.~\ref{fig:ABE_timedomain_samples}~(a)). Then, Alice and Bob can time-align their channel probing to coarsely match the IRS timing. Naturally, the ping-pong packet exchange needs to be shorter than the IRS update interval. This step may also be performed as part of the protocol initiation.

\subsubsection{Quantization}
\begin{figure}
\centering

\subfloat[]{%
        \includegraphics[width=0.45\linewidth]{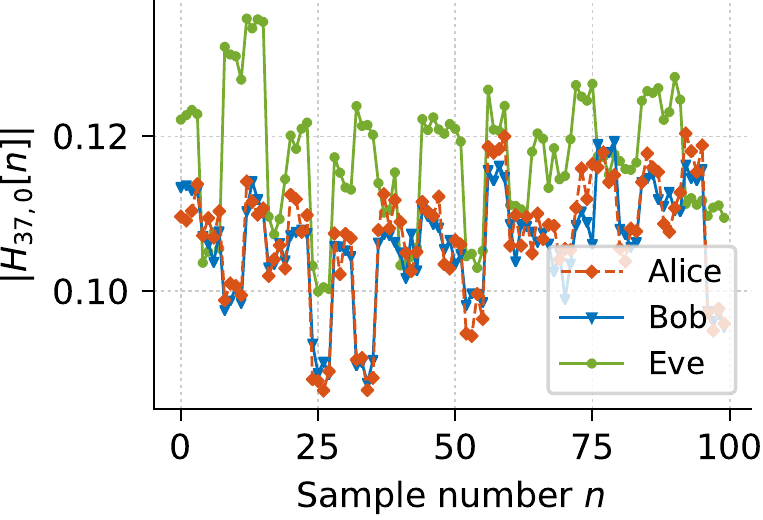}}
\hfill
\subfloat[]{%
        \includegraphics[width=0.45\linewidth]{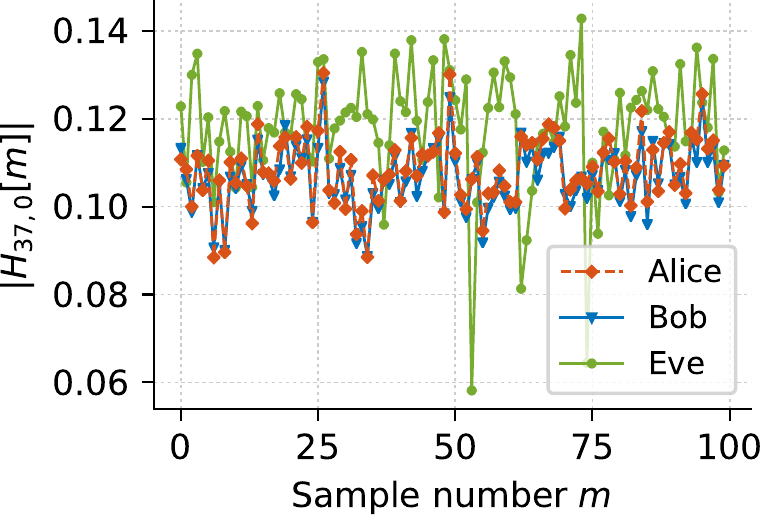}}
\caption{Time domain samples taken by Alice, Bob, and Eve on the exemplary subcarrier $k = 37$ of spatial channel $j = 0$. (a) Raw signals with $L = 4$. (b) Signals after downsampling from block averaging.}
\label{fig:ABE_timedomain_samples}
\end{figure}

A quantization stage maps Alice' and Bob's channel observations to bits to derive key material from channel measurements. As input to the quantizer, we use a series of $m$~samples for a fixed single subcarrier~$k$. Each of the $m$ samples is obtained from averaging over $L$ consecutive normalized magnitude channel estimates~(see Fig.~\ref{fig:ABE_timedomain_samples}~(b)). Due to its simplicity and equiprobable output, we here use a cumulative distribution function~(CDF)-based quantization scheme with gray coding from the literature~\cite{patwari_high-rate_2010}.

\section{Prototype Implementation}
We now describe our prototype system consisting of an IRS and commodity \mbox{Wi-Fi} transceivers.% before presenting the results of our practical evaluation in Section~\ref{sec:evaluation}.

\begin{figure}
\hspace*{\fill}%
\subfloat[]{%
        \includegraphics[width=0.35\columnwidth]{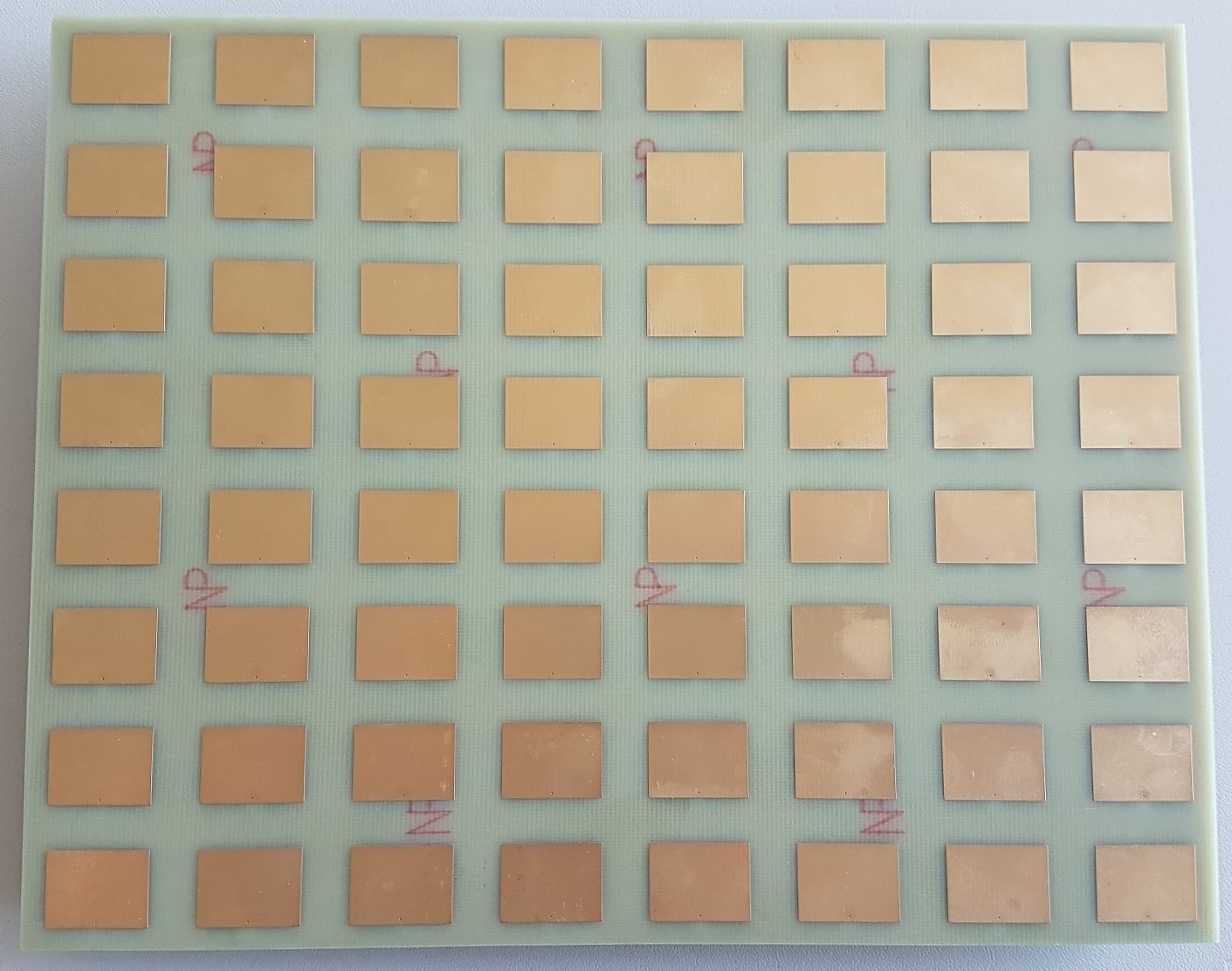}}
\hfill
\subfloat[]{%
\includegraphics[width=0.4\columnwidth]{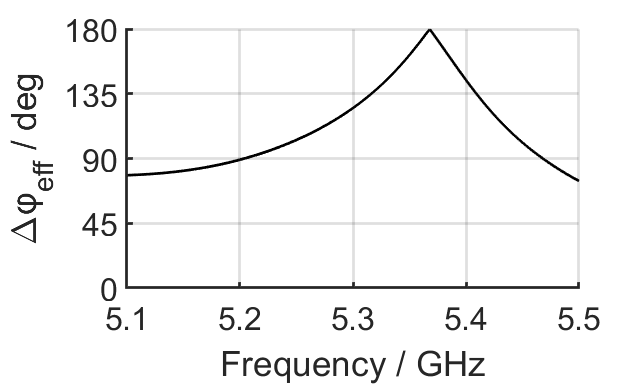}
}
\hspace*{\fill}%
\caption{Intelligent reflecting surface prototype. (a) Front view with patch elements (\SI{20}{cm} x \SI{16}{cm}). (b) Unit cell phase response.}%(b) Back view with control lines and biasing circuitry.}
\label{fig:irs_foto_phase}
\end{figure}

\subsubsection{IRS prototype}
%\begin{figure}
%\centering
%\includegraphics[width=0.49\linewidth]{figures/phase_response.png}
%\caption{Unit cell phase response.}
%\label{fig:unit_cell_phase}
%\end{figure}
We use two low-cost IRS prototypes with $64$~binary-phase tunable elements each, offering $2^{128} \approx 3.4 \times 10^{38}$ surface configurations. One IRS consists of structurally identical elements that are arranged in an $8 \times 8$~array on standard FR4 PCB substrate (see Fig.~\ref{fig:irs_foto_phase}~(a)). The elements are linearly polarized rectangular patch reflectors on top of a ground plane. The resonance frequency of the elements can be individually switched using a PIN~diode to shift the phase response. Each element has a power consumption of approximately \SI{1.5}{\mW} when the corresponding PIN diode is switched on, resulting in an average power consumption of \SI{0.75}{\mW} per element for randomized configurations. The IRS is configured by a conventional microcontroller with shift registers.%For configuration of the IRS, a conventional microcontroller with a serial interface is attached to the backside of the surface.

To obtain the maximum phase alteration, the reflection factor of the IRS is measured under the two extreme conditions when all elements are switched on and off. %The resulting phase difference between these two states is used to evaluate the array.
The measured phase difference of the IRS prototype is shown in Fig.~\ref{fig:irs_foto_phase}~(b) and is by definition limited to the range of \SI{0}{\degree} to \SI{180}{\degree}. The measurements were taken with the direction of both the incident wave and the reflected wave perpendicular to the IRS.

In our experimental setup, we synchronize the node's packet exchange and the surface configuration timing. We use the ISAAC pseudorandom number generator~(PRNG)~\cite{jenkins_isaac_1996} seeded from \texttt{/dev/random}~\cite{ubuntu_manpage_random_2020} to generate random surface configurations~$c_i[n]$. The configurations should be erased and remain secret to prevent environment reconstruction attacks.

\subsubsection{Wi-Fi NICs}
In our prototype system, each party Alice, Bob, and Eve consists of a single-board computer equipped with an ath9k-based PCIe NIC~\cite{xie_precise_2015} for IEEE 802.11n \mbox{Wi-Fi} in a 2x2 MIMO configuration, using off-the-shelf linearly polarized \mbox{Wi-Fi} antennas. The participants transmit at \SI{5}{dBm} and allocate a~\SI{40}{\MHz} wide channel at~\SI{5300}{\MHz} (\mbox{Wi-Fi} channel~$60$), close to the IRS' optimum operation frequency. During channel probing, the devices rapidly exchange packets in a ping-pong manner. For each packet and spatial MIMO channel, we obtain a complex vector containing the CSI data for each of the $114$ non-zero OFDM subcarriers. Extending (\ref{eq:channel_probing_alice}) and (\ref{eq:channel_probing_bob}), we denote them as $\hat{H}^A_{k,j}$ and $\hat{H}^B_{k,j}$, with the $j^{th}$ entry in the MIMO channel matrix.

\section{Performance Evaluation}
\label{sec:evaluation}
We now present measurement results from experiments with our prototype IRS-assisted key generation system.

\subsection{Number of Active Elements}
\label{sec:active_elements}
From (\ref{eq:h_eff}), we expect the effectiveness of channel randomization to scale with the IRS size, \ie, the number of elements~$N$. Intuitively, a large IRS increases the likelihood that a portion of the transmitted signal falls on the surface to affect the channel between Alice and Bob. Furthermore, as $N$ increases, more clearly distinctive channel states will be available.

\begin{figure}
\centering
\includegraphics[width=0.55\columnwidth]{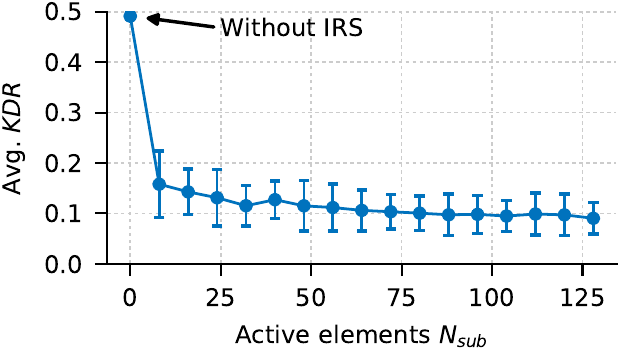}
\caption{KDR against the number of active IRS elements $N_{sub}$ with indication of mean and standard deviation.}
\label{fig:kdr_vs_N}
\end{figure}

To investigate the impact of $N$, we randomly select $N_{sub}$ elements from the surface to be used for $m=10000$ random configurations. The remaining $128 - N_{sub}$ elements are configured randomly but remain static. For all configurations, we measure the channels $\hat{H}^A_{k,j}[n]$ and $\hat{H}^B_{k,j}[n]$ with Alice and Bob \SI{3}{\m} and \SI{1.5}{\m} apart from the IRS. Then, we derive keys $K^A$ and $K^B$ for each subcarrier and spatial channel and calculate the KDR as the number of bit errors $N_e$ per $N_K$ key bits. As expected, the KDR decreases with increasing $N_{sub}$ as shown in Fig.~\ref{fig:kdr_vs_N}. Note that we have also included $N_{sub} = 0$ as a special case where the IRS remains static. Here, it is evident from the KDR of approximately 50\% that key generation is infeasible in the static environment, \ie, without the IRS.

\subsection{Distance Variation}

\begin{figure}
\centering
\includegraphics[width=0.4\columnwidth]{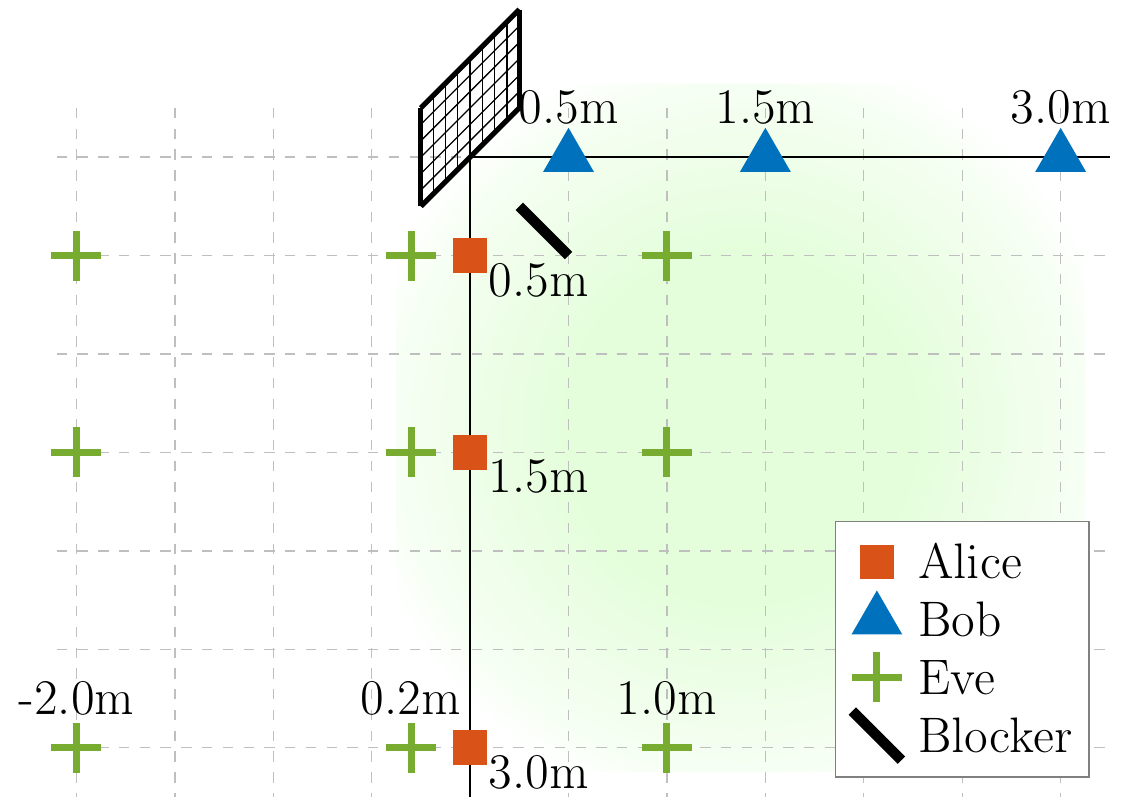}
\caption{Floorplan of the measurement setup indicating the relative positions of the parties and the approximate area of IRS radiation (green).}
\label{fig:floorplan}
\end{figure}

We evaluate IRS-assisted CRKG under varying distances of Alice and Bob to the IRS and for multiple distances of Eve to Alice. Therefore, we conduct experiments in the basement of our institute building, which is a long-term static environment. The experimental setup is illustrated in Fig.~\ref{fig:floorplan} and indicates the positions of Alice, Bob, Eve, and the IRS.

We fix Alice' distance to the IRS to \SI{3}{\m} and vary Bob's distance to the IRS between \SI{0.5}{\m}, \SI{1.5}{\m}, and \SI{3}{\m}. Then, we invert the order, fixing Bob at \SI{3}{\m} and varying Alice' distance. For each setting, we vary ($i$)~Eve's distance to Alice between \SI{0.2}{\m}, \SI{1}{\m}, and \SI{2}{\m}, ($ii$)~the number of active IRS elements between $32$, $64$, and $128$, and ($iii$)~LOS and NLOS conditions between Alice and Bob by placing a metallic blocker. For every iteration, Alice and Bob measure $m=400$~surface configurations with $L=4$.

\begin{figure}[!t]%
\centering
\subfloat[]{\includegraphics[width=0.5\columnwidth]{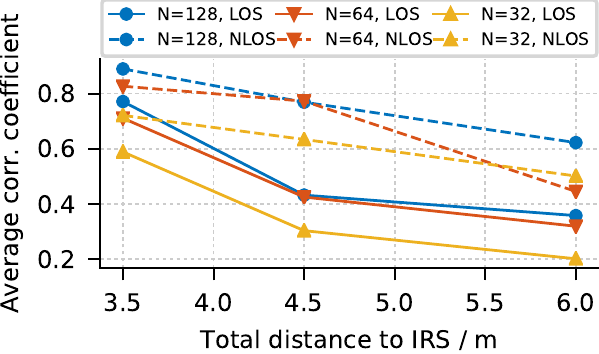}}
\hfill
\subfloat[]{\includegraphics[width=0.5\columnwidth]{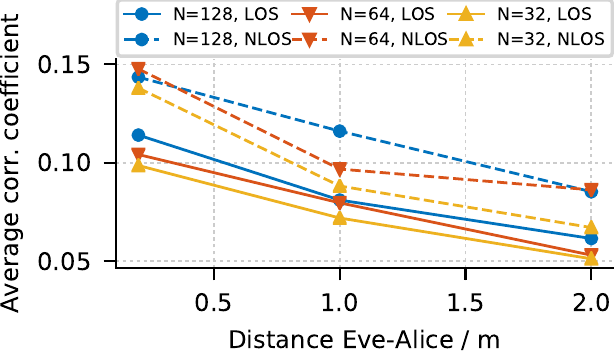}}
\caption{(a) Average correlation coefficient of signals received by Alice and Bob for varying distances to the IRS. 
(b) Average correlation coefficient between signals gathered by Alice end Eve for varying positions of Eve relative to Alice.}
\label{fig:distance_experiments}
\end{figure}

To assess the IRS-induced channel variations, 
we calculate the average of the absolute Pearson correlation coefficients $\rho^{A,B}_{k,j}$ of Alice' and Bob's magnitude channel measurements across MIMO channels and subcarriers. We plot the results over the total node distance to the IRS for $32$, $64$, and $128$ active elements and for LOS and NLOS conditions in Fig.~\ref{fig:distance_experiments}~(a). Here, increasing the distance to the IRS reduces the IRS impact, as the loss of the channels to the IRS, $h_{i,k}$ and $g_{i,k}$, increases. Further, in accordance to the previous experiment, more surface elements help to increase the channel variation. Further, the surface impact rises for NLOS channel conditions as the surface-independent direct component $d_k$ in (\ref{eq:h_eff}) is reduced.

In the same manner as before, we calculate the average of the absolute Pearson correlation coefficients $\rho^{A,E}_{k,j}$ between Eve's and Alice' channel observations, made from the signals transmitted by Bob. 
We plot the results over Eve's distance in Fig.~\ref{fig:distance_experiments}~(b) for LOS and NLOS scenarios with varying number of active surface elements~$N$. As expected and in accordance with spatial decorrelation properties~\cite{goldsmith_wireless_2005}, the correlation of Eve's channel observations reduces with distance.

\begin{figure}
\centering
\includegraphics[width=0.6\columnwidth]{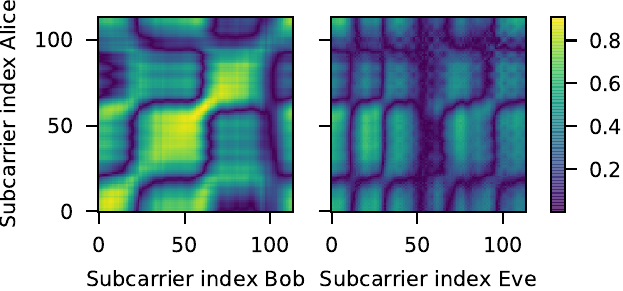}
\caption{Pearson correlation coefficient matrices of Alice and Bob (left) and Alice and Eve (right).}
\label{fig:covheat}
\end{figure}

Fig.~\ref{fig:covheat} shows the two-dimensional Pearson correlation matrices of Alice, Bob, and Eve for exemplary spatial channels, illustrating covariance relationships across the parties and subcarriers. In the experiment, Eve was at \SI{0.2}{\m} distance from Alice. As expected, Alice and Bob~(left) make strongly dependent observations on the same subcarriers, as indicated by the diagonal matrix entries. The off-diagonal components reveal that care should be taken when using frequency diversity for key generation, since some subcarriers seem to share a certain covariance. Fig.~\ref{fig:covheat}~(right) shows the correlation of Eve's observations which are mostly uncorrelated, however, medium correlation values occur occasionally.

\subsection{Rates}
\label{sec:rates}

The KGR is obtained by $\frac{N_K}{T_K}$, where $N_K$ is the number of key bits obtained per time interval $T_K$. In the context of this work, we measure the KGR at the output of the CDF quantizer. 

Note that the channel changes at the rate the IRS configuration is updated, allowing to boost key generation rates. That is, in contrast to traditional channel models, the IRS allows immediate switching of channel characteristics. Hence, IRS-assisted CRKG is not bounded by the classical channel coherence time to achieve sufficient randomness. Assuming extraction of a single bit from each IRS configuration, the KGR is upper bounded by $(T_{su} + L\, T_p)^{-1}$, where $T_{su}$ is the surface update time, $T_p$ is the channel probing interval, and $L$ is the oversampling factor. 

In our implementation, a bi-directional packet exchange $T_p$ takes approx. \SI{2}{\ms} and the surface update time $T_{su}$ is approx. \SI{2}{\ms}, which can be further reduced through technical optimization. From the measurement data used in Section~\ref{sec:active_elements} with $N=128$, we calculate the KGR and KDR shown in Table~\ref{tab:rates}. The results for varying oversampling factors~$L$ at two exemplary subcarriers demonstrate that a trade-off between measurement time and KDR reduction is possible. 
We emphasize that only a single subcarrier from a single spatial channel is used and therefore we expect that the KGR can easily be further increased. 
The residual KDR of Alice and Bob stems from the CDF quantizer, when input values are close to the quantizer thresholds. 

\begin{table}[ht]
\sisetup{round-mode=places}
\sisetup{group-separator = {}}
\centering
\captionsetup{justification=centering}
\caption{Single subcarrier KGR and KDR}
\label{tab:rates}
\begin{tabular}{@{}rccccc@{}}
\toprule

%%%%%%% new results from improved surface controller
                    & \textbf{$L=1$}    & \textbf{$L=2$} & \textbf{$L=3$} & \textbf{$L=4$}\\
\toprule
\textbf{KGR [bit/s]}                    & 237.45   & 160.58   & 121.30  &  97.39 \\
\textbf{KDR A/B, $k=25$}                    & 0.260   & 0.157   & 0.108  & 0.083  \\
\textbf{KDR A/B, $k=94$}                    & 0.218   & 0.120  & 0.082  &  0.065 \\

\arrayrulecolor{black}\bottomrule 
\end{tabular}
\end{table}

\subsection{Randomness}
Using the NIST's statistical test suite for random number generators~\cite{rukhin_statistical_2010}, we assess the randomness of binary sequences generated from the prototype IRS-assisted CRKG system on a single subcarrier with $L=4$ to cover $m=$~\SI{300000}~surface configurations. In the experiments, Alice and Bob are located \SI{0.5}{m} and \SI{3}{m} away from the IRS, respectively, (C, D), and in a metallic shielding box together with the IRS~(A, B). Using the CDF quantizer~\cite{patwari_high-rate_2010} with \SI{1}{\bit} and \SI{2}{\bit} resolution, we produce binary sequences of lengths \SI{300000}~(A, C) and \SI{600000}{bits}~(B, D). We apply the tests that are applicable to the given sequence lengths and list the results in Table~\ref{tab:nist_results}, showing that all sequences pass the tests.

\begin{table}[ht]

\sisetup{round-mode=places}
\sisetup{group-separator = {}}
\centering
\caption{NIST statistical test suite $p$-value results~\cite{rukhin_statistical_2010}.}
\label{tab:nist_results}
\begin{tabular}{@{}rccccc@{}}
\toprule
                    & \textbf{A}    & \textbf{B} & \textbf{C} & \textbf{D}\\
\toprule
\textbf{Frequency}                          & 1.00000   & 1.00000   & 1.00000  &  1.00000 \\
\textbf{Block Frequency}                    & 0.86108   & 0.64011   & 0.28217  &  0.83365 \\
\textbf{Runs}                               & 0.85513   & 0.99588   & 0.28984  &  0.26890 \\
\textbf{Longest Runs}                       & 0.77137   & 0.32486   & 0.34298  &  0.07352 \\
\textbf{Binary Matrix Rank}                 & 0.09973   & 0.04950   & 0.09466  &  0.68954 \\
\textbf{DFT}                                & 0.69998   & 0.85895   & 0.93210  &  0.79094 \\
\textbf{\begin{tabular}[c]{@{}r@{}}Non-overl.\\Templ. Matching\end{tabular}}         & 0.02935   & 0.46983   & 0.87188  &  0.48896 \\
\textbf{Universal}                          & -         & 0.13809   & -        & 0.133209  \\
\textbf{Serial}                             & \tline[c]{0.59214\\0.88924}    & \tline[c]{0.54280\\0.78167}  &  \tline[c]{0.34804\\0.32932} & \tline[c]{0.73928\\0.76001} \\
\textbf{Approx. Entropy}                    & 0.81847   & 0.12792   & 0.403202  & 0.16795 \\
\textbf{Cum. sums (Fwd)}                    & 0.12669   & 0.27253   & 0.190075  & 0.22361 \\
\textbf{Cum. sums (Rev)}                    & 0.12669   & 0.27253   & 0.190075  & 0.22361 \\

\arrayrulecolor{black}\bottomrule 
\end{tabular}

\end{table}

\section{Discussion and Future Work}

Our prototype system with commodity \mbox{Wi-Fi} transceivers demonstrates that conventional radios can successfully perform wireless key generation in static environments complemented by an IRS. Although utilizing a MIMO system, we considered single-subcarrier signals, showing that the concept is applicable to resource-constrained IoT devices with single carrier radios as well. However, larger IRS deployments are needed for a sufficient impact on coarse channel measurements such as RSS. An actively cooperating IRS for key generation requires a trust model between devices and their environment. However, an IRS could alternatively also randomize the channel continuously without participation in the CRKG protocol. 

Building on our prototype system, future work will investigate how the IRS could improve the channel conditions of Alice and Bob while providing randomness for key generation. Also, future work could investigate requirements to enhance channel randomization, \eg, from spatial and frequency diversity, the number, size, and placement of IRS elements, the surface modulation signal, and inter-element correlations. More work is needed to evaluate the approach in non-static environments. Finally, we stress the need for a sound security analysis of IRS-assisted CRKG.

\section{Conclusion}

In this paper, we outline a novel wireless key generation system assisted by an IRS, \ie, a smart radio environment. Here, we leverage a time-dependent randomization of the IRS configuration to overcome static radio environments which are not suited for wireless key generation. 
The passive IRS provides an entropy source, allowing arbitrary near-by users to perform key generation. The proposed system works with conventional CRKG implementations, completely eliminating special measures for static environments at the user terminals, \eg, complex hardware extensions. Demonstrating the practical feasibility to establish cryptographic key material, we have implemented a functional prototype system using commodity \mbox{Wi-Fi} transceivers and a low-cost IRS prototype.

\section*{Acknowledgements}
This work was partially funded by the German Federal Ministry of Education and Research~(BMBF) (Grant 16KIS1234K MetaSEC) and by the German Research Foundation~(DFG) within the framework of the Excellence Strategy of the Federal Government and the States - EXC2092 CASA - 390781972.

\bibliographystyle{IEEEtranS}
% argument is your BibTeX string definitions and bibliography database(s)
\small
\bibliography{metasec_refs}

% that's all folks
\end{document}